# Asymmetric Graphene Model Applied to Graphite-like Carbon-based Ferromagnetism

Norio Ota*, Narjes Gorjizadeh** and Yoshiyuki Kawazoe**

*R&D Division, Hitachi Maxell Ltd., 1-1-88 Ushitora, Ibaraki-city, Osaka 567-8567 JAPAN

** Institute for Materials Research, Tohoku Univ., 2-1-1 Katahira, Aoba-ku, Sendai 980-8577 JAPAN

Several experiments have recently found room-temperature ferromagnetism in graphite-like carbon based materials. This paper offers a model explaining such ferromagnetism by using an asymmetric nano-graphene. Our first typical model is $C_{48}H_{24}$ graphene molecule, which has three dihydrogenated (-$CH_2$) zigzag edges. There are several multiple spin states competing for stable minimum energy in the same atomic topology. Both molecular orbital and density function theory methods indicate that the quartet state(S=3/2) is more stable than that of doublet (S=1/2), which means that larger saturation magnetization will be achieved. We also enhanced this molecule to an infinite length ribbon having many (–$CH_2$) edges. Similar results were obtained where the highest spin state was more stable than lower spin state. In contrast, a nitrogen substituted (-NH) molecule $C_{45}N_3H_{21}$ demonstrated opposite results. that is, the lowest spin state(S=1/2) is more stable than that of highest one(S=3/2), which arises from the slight change in atom position.

## 1. Introduction

Carbon based room-temperature ferromagnetism is a long term research target for both science and industry[1]. Carbon molecules associated with light elements like hydrogen, nitrogen and/or oxygen is very attractive in view of ultra light organic magnet. It will be a key material for future ecological magnetic material and new spintronics application. Recently, several experiments show room temperature ferromagnetism in modified graphite materials. P.Esquinazi et al.[2] predicted ferromagnetic magnetization loop in proton ion irradiated highly oriented pyrolytic graphite (HOPG). Magnetic moments with orders of $5 \times 10^{-6}$ emu was observed at room-temperature having a coercive force of 100 Oe. Saitama university group opened impressive photograph and data[3,4]. Synthesized powder is magnetically attracted by a permanent magnet. Estimated Curie temperature is very high up to 800K. Starting material is tri-ethylamine following direct pyrolysis at 925C resulting 0.5emu/g saturation magnetization and graphite like X ray diffraction pattern. Very recent experiment in 2009 is opened by Y.Wang et al[5] entitled "Room-Temperature Ferromagnetism of Graphene".
Their starting material is graphite oxide(GO) made by Hummers method[6]. Separated GO was strongly reduced using hydrazine and annealed at 400-600C. Saturation magnetization is 0.02emu/g at 300K, whereas coercive force 40 Oe. Those experiments encourage us to open a new door to carbon magnets. However, suitable models and guiding principles how to design those materials are not clear yet. This paper firstly offers a model using asymmetric graphene as like $C_{48}H_{24}$ molecule, infinite length ribbon and nitrogen substituted $C_{45}N_3H_{21}$ molecule.

These five years, another amazing discovery was single sheet graphene firstly reported by K.Novoselov et al.[7]. Extraordinary electronic transport properties nominate graphene as a post silicon[8,9], also extraordinary mechanical strength as a post iron steel[10].

There were many theoretical predictions on $\pi$-electron network carbon systems. Especially, zigzag edge graphene magnetism is very exciting in view of localized density states near Fermi energy[11-14]. Focusing on infinite length graphene ribbon, M.Fujita et al[11] indicated an anti-ferromagnetic feature with total magnetization zero using a tight binding model. Also, rectangular shaped graphene has been extensively studied[15-18]. The singlet state (zero magnetization) arises by both side zigzag edge modifications with various species like –H,-F,-O,-OH, and –$CH_3$.[15] Whereas, K.Kusakabe and M.Maruyama[19,20] proposed ferrimagnetic graphene ribbon model with non-zero total magnetization which has dihydrogenated carbon at one side zigzag edge, remaining opposite side mono-hydrogenated.

In this paper, we firstly report nano meter size asymmetric graphene molecule model. The reason is that there are several spin states in the same molecule topology which compete each other to get energy minimum. In previous papers[19,20] applied to infinite length ribbon, density function theory(DFT) based calculation gave strong magnetic behavior. However, a unit cell has single dihydrogenated carbon($CH_2$-) edge. In such a case, edge spin show all up or all down by a repetition nature of periodic boundary condition. We should calculate energy stability between multiple spin states of a molecule having several edges.

Two calculation methods, a semi-empirical molecular orbital (MO) method and a first principle density function theory (DFT) method are applied. MO has an advantage to use room-temperature experimentally parameterized basis functions, whereas DFT gives us more detailed spin configurations than MO at a ground state. We need to compare such two methods.

In this paper we like to apply to typical molecule $C_{48}H_{24}$ and its extension to infinite length ribbon and nitrogen substituted molecule $C_{45}N_3H_{21}$. Former $CH_2$ modified case shows stable high spin (S=3/2) state than low spin (S=1/2) one. On the contrary, nitrogen substituted (-NH) case shows opposite result, that is, low spin state is more stable than high spin state. Such remarkable difference may arise from a slight atom position change in the same molecule.

## 2. Piled Graphene Model

Proton ion irradiation experiment[1] inspired us to make a proper molecule model. Fig.1 shows piled sheets of graphene. Each element of HOPG is thought as one leaf of graphene molecule. Here, as a typical molecule model, we illustrate $C_{48}H_{21}$. All edges are mono-hydrogenated. Protons are irradiated on such piled graphenes. As imagined in Fig.1, by a shadow effect of upper molecule, only one side edge carbons are attacked by protons to modify to dihydrogenated carbons, whereas the other side remains mono-hydrogenated. Typical asymmetric molecule $C_{48}H_{24}$ will be created as illustrated in Fig.2, which is our first typical calculation model.

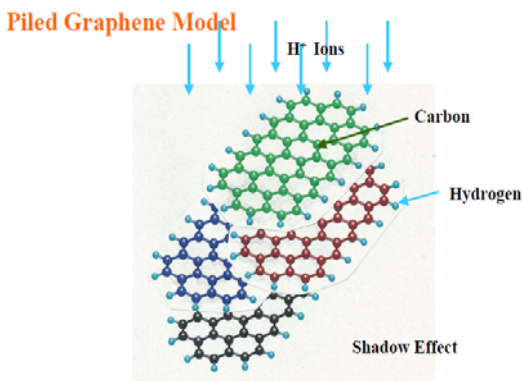

Fig.1 Graphene molecules are piled one by one, where proton ions are irradiated from top. One side of edge carbon is dihydrogenated by proton irradiation, the other side with shadowed edge carbons remains mono-hydrogenated. Such a view gives us an asymmetric graphene model.

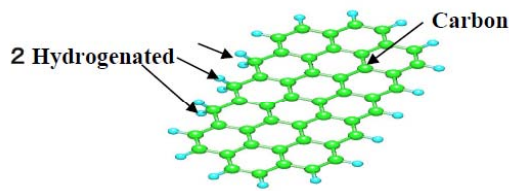

Fig.2 Asymmetric graphene molecule of $C_{48}H_{24}$. Left side of zigzag edge has three carbons that are dihydrogenated, which is a model used for calculating with both on molecular orbital (MO) and density function theory (DFT) methods.

## 3. Calculation Methods

In order to clarify magnetism of molecule, we have to obtain (i) spin density mapping, (ii) square spin momentum= S(S+1) and (iii) total molecular energy to decide which magnetic state is stable. Two calculation methods are applied. One is a semi-empirical molecular orbital method (MO) employing PM5 hamiltonian[21] in programs of Fujitsu Scigress MO V1 Pro [22] Advantage of this calculation is that basis are parameterized by many room temperature experiments, whereas some consideration is necessary because of an insufficient electron-electron correlation so called spin contamination. Another approach is to use the first principle density function theory (DFT) based method[23)24]. Our choice for this work is a hybrid density functional in the unrestricted UB3LYP[25)-27] with basis set of 6-31G[28)29]. Other generalized gradient approximation under the scheme of DFT gave us similar result of energy stability, so in this paper we typically show UB3LYP results. Non-local and semi-local hybrid integrals in UB3LYP are expected to reduce spin contamination than MO and also than conventional localized spin density approximation[30)31]. Precise energy calculation is done in 10E-8 accuracy as total energy of one molecule. Atom position accuracy is required to be the threshold of 0.00018nm. We should mind that DFT is based on zero temperature ground state. We need those two MO and DFT-UB3LYP calculation to estimate actual magnetic behavior at a room temperature.

## 4. Dihydrogenated Asymmetric Graphene Molecule

### 4.1 Molecular Orbital Calculation

It is well known that carbon atom has two bonding configurations, one is $sp^3$ tetrahedral arrangement, another is $sp^2$ arrangement. When dihydrogenation occurs at edge carbon, $sp^3$ configuration is favored remaining no spins on edge carbon. Whereas, mono-hydrogenation case, $\pi$-electron remains as an isolated spin. This basic difference affects to spin configuration and total behavior of molecular magnetism.

Fig.3 shows a spin density mapping of $C_{48}H_{24}$ obtained by MO method. Up spin (dark gray: red in color) and down spin (light gray: blue) are both peanuts like figures reflecting $\pi$-orbital. This is contour surface map at 0.001 $\mu_B/A^3$. In case of z-axis total magnetization Ms is $3\mu_B$, where $\mu_B$ is Bohr Magneton, there appears quartet state(S=3/2) as illustrated in Fig.3(a). Dihydrogenated three carbons(green) has no spins at all. In a whole molecule, there appears up and down spins simultaneously. We also obtain Ms=$1\mu_B$ spin distribution in (b) as doublet state(S=1/2).

In order to decide a stable spin state, the difference of total molecular energy is calculated.

Their difference between $3\mu_B$ and $1\mu_B$ is,

$$E(3\mu_B) - E(1\mu_B) = -16.7 \text{Kcal/mol}$$

Such minus value implies that high spin state is more stable than low spin state, in other words, larger magnetization will be realized in a molecule. We can understand why low spin state needs high energy by looking a spin distribution in (b). Inside of molecule, there is a (down-down) spin pair, which generates higher exchange integral energy.

MO method gives us a simple understanding of magnetic behavior at a room temperature, but unfortunately it gives us an insufficient spin configuration. For example, S(S+1) of quartet state is overestimated as 5.5. Reasonable one is around 3.75 from S value of 3/2.

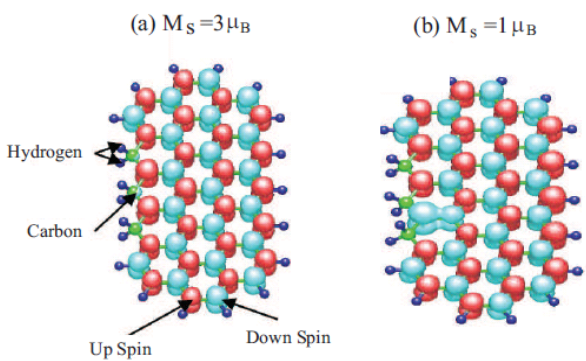

Fig.3 Spin-density contour surface of $C_{48}H_{24}$ molecule by MO method. Dark gray (red in color) indicates up spin, while light gray (blue) indicates down spin. Figure 3(a) predicts quartet state (S=3/2) while (b) predicts doublet(S=1/2). There are no spins on three dihydrogenated carbon atoms in either cases. In quartet state(a), there appears up and down spins simultaneously, whereas irregular down-down pair appears in a doublet-state(b).

### 4.2 Density Function Theory Based Analysis

In order to have more accurate electron-electron correlation and to reduce spin contamination, we applied DFT based UB3LYP calculation to $C_{48}H_{24}$ asymmetric molecule. There appears quartet state (Ms=$3\mu_B$) spin-density map in Fig.4(a). Up and down spins are alternatively arrange one by one at a contour surface of 0.001 $\mu_B/A^3$. Whereas in case of doublet state (Ms= $1\mu_B$) as shown in Fig.4(b), we can see complex spin arrangement inside of molecule as like up-up spin pairs and down-down pairs. As summarized in Table 1, S(S+1) value of quartet state is 3.76 which is close to simple estimation of 3.75 from S=3/2, while that of doublet state is 0.96 almost close to 0.75 for S=1/2.

Molecular size is completely same as 0.924nm by 1.679nm in both magnetic states summarized in Table 1. Also at a zigzag edge center carbon, carbon-to-carbon distance is the same as 0.149nm, angle of (C-C-C) is the same as 115.3 degree. Those shows the same atomic arrangement between quartet and doublet states.

Only the difference between two states is calculated as the total energy difference caused by electron-electron correlation as like.

$$E(3\mu_B) - E(1\mu_B) = -7.9 \text{ Kcal/mol}.$$

Again, larger magnetic moment (quartet state) is more stable than smaller one(doublet state), Comparing with MO result, the energy difference is almost half by employing a DFT method.

From such energy difference, we can estimate temperature stability. Above value of 8 Kcal is translated to be 3900K as a thermal excitation temperature of a molecule, which predicts stable high spin state at a room temperature.

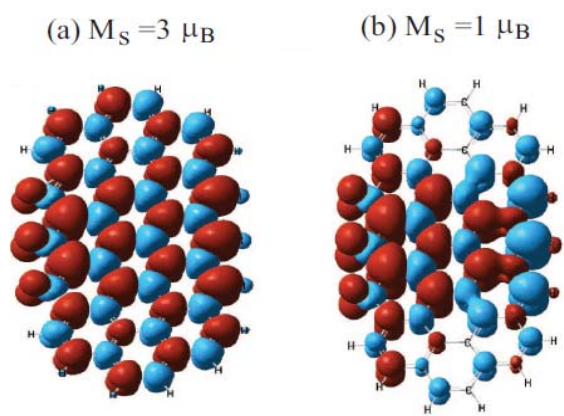

Fig.4 Spin density of (a)quartet state and (b)doublet state calculated by DFT-UB3LYP at contour surface of 0.001 $\mu_B/A^3$. In (a), there appears regularly aligned up-down pairs, whereas in (b) up-up and down-down complex pairs.

TABLE I: Summary of major calculated results by a Gaussian DFT program using UB3LYP method with 6-31G basis.

| | $C_{48}H_{21}$ | $C_{48}H_{24}$ | | $C_{45}N_3H_{21}$ | |
|---|---|---|---|---|---|
| Spin State | Singlet | Quartet | Doublet | Quartet | Doublet |
| S(S+1) by DFT calculation | 0.0 | 3.76 | 0.96 | 3.75 | 0.75 |
| Energy Difference * | — | -7.90 Kcal/mol | | +11.64 Kcal/mol | |
| Molecular size (width × length) | 0.928 nm × 1.663 nm | 0.924 × 1.679 | 0.924 × 1.679 | 0.917 × 1.665 | 0.913 × 1.663 |
| C-C, C-N bond length ** | 0.140 nm | 0.149 | 0.149 | 0.141 | 0.139 |
| C-C-C, C-N-C bond angle ** | 122.0 degree | 115.3 | 115.3 | 118.9 | 122.0 |

* Energy difference = E (Quartet) - E (Doublet)
** Bond length, bond angle are at zigzag edge center atom

The energy spectra of quartet and doublet states for spin up and spin down obtained from DFT-UB3LYP calculations are depicted in Fig.5 In this figure, solid lines are electrons filled states in the ground state, while the dotted lines are empty. Quartet state is more stable than doublet one by comparing the energy gap between highest molecular orbital (HOMO) and lowest unoccupied molecular orbital (LUMO).

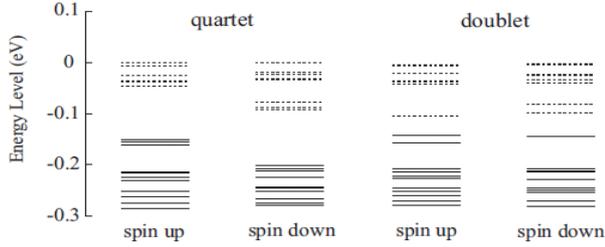

Fig.5 Energy spectra of quartet and doublet states of $C_{48}H_{24}$ for up and down spins, obtained from DFT-UB3LYP calculations. Solid lines are electron-filled states, while dotted lines are empty.

### 5, Infinite Length Asymmetric Ribbon

In above session, the essence of our result is that high spin state(quartet state) is more stable than low spin one(doublet state) in energy comparison To simulate the experimental situation of proton irradiated HOPG, we should enhance the molecular size from this minimal molecule to larger one. There may be many different length graphene molecules in actual HOPG. Extreme maximum is an infinite length ribbon. To compare multiple spin states, we applied larger unit cell employing three dihydrogenated carbons. Result is shown in Fig.6. This image has three neighboring super cells, where only the center cell sketched spin densities. The highest spin state is shown in figure 6(a) having more stable energy level comparing than lower spin state in (b). High spin state is energetically more stable as 5.4Kcal/unit cell than low one. From this result, we can expect simple extrapolation that larger magnetization will be realized from nano to macro scale in $CH_2$- modified asymmetric graphene.

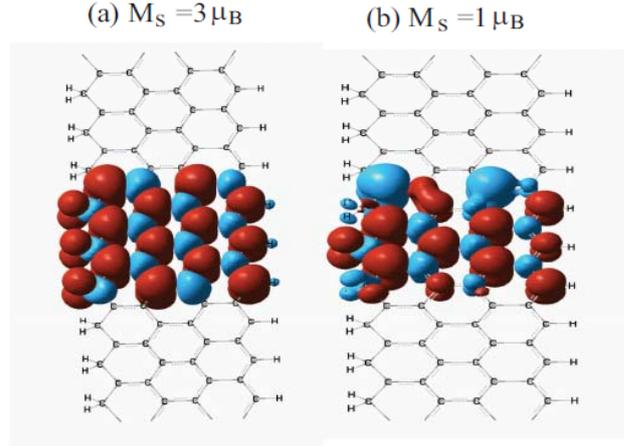

Fig.6 Spin density of infinite length asymmetric graphene ribbon. One unit cell for periodic calculation is shown as colored region having three $CH_2$- zigzag edges. This image has three neighboring super cells, where only the center cell sketched spin densities.

### 6, Nitrogen Substituted Molecule

Common understanding of magnetism is that high spin state is not stable than lower one. However, above mentioned dihydrogenated asymmetric graphene shows irregular result. Here, we like to check another type of asymmetric graphene. Triggered by an experiment of Reference 2)、 we tried to substitute three edge carbons by three nitrogen atoms as a molecule $C_{45}N_3H_{21}$ shown in Fig.7. All nitrogen is mono-hydrogenated (NH-) respectively. Nitrogen has one lone pair electrons (up and down spin pair in a same $\pi$ orbit). This brings a disappearance of local net moment,which is a similar situation with $CH_2$- case. Fig.7 is a result for spin density contour surface at $0.01 \mu_B/A^3$ ,where (a) is for quartet state and (b) for doublet one.
Energy difference between two magnetic states is,
$$E(3\mu_B) - E(1\mu_B) = +11.6 Kcal/mol$$
that is, doublet state (Ms=$1\mu_B$) is more stable than quartet one. This is an opposite result with that of $CH_2$- modified molecule.

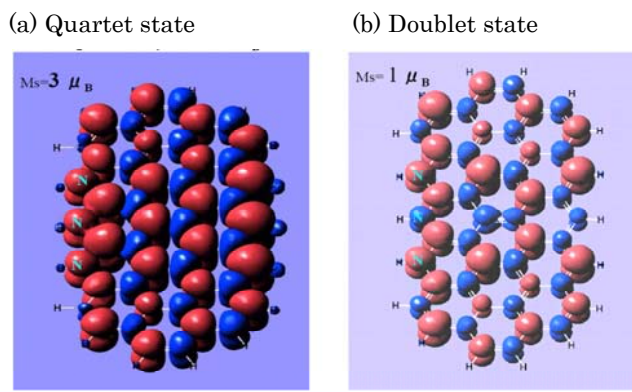

Fig.7 Three nitrogen atoms substituting for left side zigzag edge carbon atoms to form $C_{45}N_3H_{21}$. (a)Quartet and (b)doublet state are seen at $0.01 \mu_B/Å^3$ spin contour surface. Molecular size and carbon-to-carbon distances are slightly different as summarized in Table1 resulting that the doublet state is more stable than the quartet.

To understand this result, we compared detailed data as shown in table 1. Square S(S+1) values are 3.75 and 0.75 for quartet and doublet cases, which are reasonable values considering S=3/2 and 1/2 respectively. Molecular size is a little bit smaller in doublet one, also shorter carbon-to-carbon length. Angle of (C-N-C) at substituted nitrogen is 122.0 degree for doublet and 118.9 for quartet. Those results teach us that the magnetism of asymmetric graphene should be carefully treated considering modified chemical elements and atomic position change. In other words, we should notice that there occurs slight atomic position change depend on the spin state even in the same molecule topology, which finally brings total energy difference.

## 7, Conclusion

Recent several experiments indicate room-temperature ferromagnetism of graphite like materials. This paper firstly offers a nano meter size asymmetric graphene model to explain such ferromagnetism by introducing three typicals as $CH_2$-bonded $C_{48}H_{24}$ molecule, infinite length ribbon and CH-bonded $C_{45}N_3H_{21}$ molecule. There are several multiple spin states and compete each other to get stable minimum energy under the same atomic topology. We concluded as,

(1) In $C_{48}H_{24}$ molecule, quartet state(Ms=$3\mu_B$) is more stable than doublet ($1\mu_B$) in both calculations of a molecular orbital method (MO) and a first principle density function theory(DFT) method. Energy difference between two states is 8Kcal/mol in DFT calculation, which means to be tolerable against 3900K thermal excitation. This predicts that larger magnetization is more stable even at room temperature.

(2) Extending graphene length to infinite, periodic boundary condition calculation is applied to $CH_2$-bonded graphene ribbon. One super cell has three $CH_2$-zigzag edges to compare stable spin states. Again, high spin state is stable as like molecule. By those results from nano to macro scale, $CH_2$- asymmetric graphene show larger magnetization expecting ferromagnetism.

(3) Nitrogen substituted asymmetric molecule $C_{45}N_3H_{21}$ shows that doublet state is more stable than quartet one, which is opposite result with $CH_2$-bonded case. There appears slight atom position change between two spin states. Magnetism of asymmetric graphene should be carefully treated considering modified chemical elements and atomic position change.

By those results, we can believe that $CH_2$- modified asymmetric graphene realize ferromagnetism at room-temperature.

## Acknowledgements

Narjes Gorjizadeh would like to thank the crew of the Center for Computational Materials Science, Institute for Materials Research of Tohoku University for their support of the Hitachi SR11000(model K2) supercomputer system, and Global COE Program "Materials Integration (International Center of Education and Research),Tohoku University," MEXT, Japan, for financial support.